\documentclass[twocolumn,preprintnumbers,amsmath,amssymb]{revtex4}  

\usepackage{graphicx}
\usepackage{dcolumn}
\usepackage{bm}
\usepackage{multirow}
\usepackage{color}
\usepackage{amsmath}

\usepackage[colorlinks]{hyperref}

\begin{document}
\title{Magneto-optics in transition metal diselenide monolayers }

\author{G.~Wang$^1$}
\author{L.~Bouet$^1$}
\author{M.~M.~Glazov$^2$}
\author{T.~Amand$^1$}
\author{E.~L.~Ivchenko$^2$}
\author{E.~Palleau$^1$}
\author{X.~Marie$^1$}
\author{B.~Urbaszek$^1$}
\affiliation{%
$^1$ Universit\'e de Toulouse, INSA-CNRS-UPS, LPCNO, 135 Avenue de Rangueil, 31077 Toulouse, France}
\affiliation{%
$^2$ Ioffe Institute, 194021 St.-Petersburg, Russia}

\begin{abstract}
We perform photoluminescence experiments at 4~K on two different transition metal diselenide monolayers, namely MoSe$_2$ and WSe$_2$ in magnetic fields $B_z$ up to 9~T applied perpendicular to the sample plane. In MoSe$_2$ monolayers the valley polarization of the neutral and the charged exciton (trion) can be tuned by the magnetic field, independent of the excitation laser polarization. In the investigated WSe$_2$ monolayer sample the evolution of the trion valley polarization depends both on the applied magnetic field and the excitation laser helicity, while the neutral exciton valley polarization depends only on the latter. Remarkably we observe a reversal of the sign of the trion polarization between WSe$_2$ and MoSe$_2$. For both systems we observe a clear Zeeman splitting for the neutral exciton and the trion of about $\pm2$~meV at $B_z\mp9$~T. The extracted Land\'{e}-factors for both exciton complexes in both materials are $g\approx -4$.  
\end{abstract}


\maketitle

\section{Introduction}
\label{sec:intro}
The two-dimensional (2D) semiconductor structures based on monolayer (ML) transition metal dichalcogenides (TMDCs) show very strong light-matter interaction, with an absorption of the order of 10\% per ML in the visible region of the optical spectrum \cite{Mak:2010a,Splendiani:2010a,Zhang:2014a}. This strong interaction can be explored in microcavity physics and for laser and light emitting diode (LED) devices. The optical properties of ML TMDCs are governed by strongly bound excitons (binding energy is of the order of $0.5$~eV) \cite{Tawinan:2012a,Komsa:2012a,Ross:2013a,He:2014a,Ugeda:2014a,Chernikov:2014a,Ye:2014a,Wang:2015b,Klots:2014a}.  The absence of an inversion centre in the lattice of  TMDC  MLs  together with the strong spin-orbit (SO) interaction in these materials leads to a coupling of carrier spin and $\bm k$-space valley dynamics. As a result the circular polarization ($\sigma^+$ or $\sigma^-$) of the absorbed or emitted photon can be directly associated with selective exciton generation in one of the two non-equivalent $K$-valleys: $K^+$ or $K^-$, respectively \cite{Xiao:2012a, Cao:2012a,Mak:2012a,Sallen:2012a,Kioseoglou:2012a,Jones:2013a,Mak:2014a}. Excitonic resonances do not only dominate single- and two-photon absorption, but also strongly influence second harmonic generation (SHG) \cite{Wang:2015b} and interactions with plasmons \cite{Najmaei:2014a}. The exact symmetry, degeneracy and energy spacing of the exciton levels is still under debate. To address the spin- and valley-dependent fine structure of excitonic levels we perform magneto-optical experiments on ML TMDCs, where the magnetic field lifts the underlying degeneracies and extract the effective Land\'e factors, a technique successfully used in the past to study the bandstructure and excitonic effects of semiconductors \cite{Ivchenko:1995a}.\\
\indent Very recently lifting of the $K$-valley degeneracy through the application of a magnetic field perpendicular to the monolayer plane has been reported for MoSe$_2$ \cite{Macneill:2015a,Li:2014a} and WSe$_2$ \cite{Aivazian:2015a,Srivastava:2015a}. An energy splitting $\Delta_Z$ between the $\sigma^+$ and  $\sigma^-$ polarized photoluminescence (PL) components has been reported, which correspond to inter-band recombination in the $K^+$- and $K^-$-valley, respectively~\cite{Xiao:2012a,Cao:2012a,Mak:2012a,Zeng:2012a,Sallen:2012a,Kioseoglou:2012a}. The splitting  $\Delta_Z$  increases roughly linearly with the applied magnetic field, although the values reported for ML WSe$_2$ differ by a factor of up to 3 \cite{Aivazian:2015a,Srivastava:2015a}.
The aim of this comparative study is to perform experiments on MoSe$_2$ and WSe$_2$ monolayers in magnetic fields up to $|B_z| = 9$~Tesla in the same set-up, and to measure the neutral exciton and trion emission energy and polarization. We vary the laser excitation energy and helicity, which allows us to distinguish between the laser induced and magnetic field induced valley polarization, which show very different dependencies on the applied magnetic field when comparing MoSe$_2$ with WSe$_2$ monolayers. 

\begin{figure*}[t]
\includegraphics[width=0.95\textwidth]{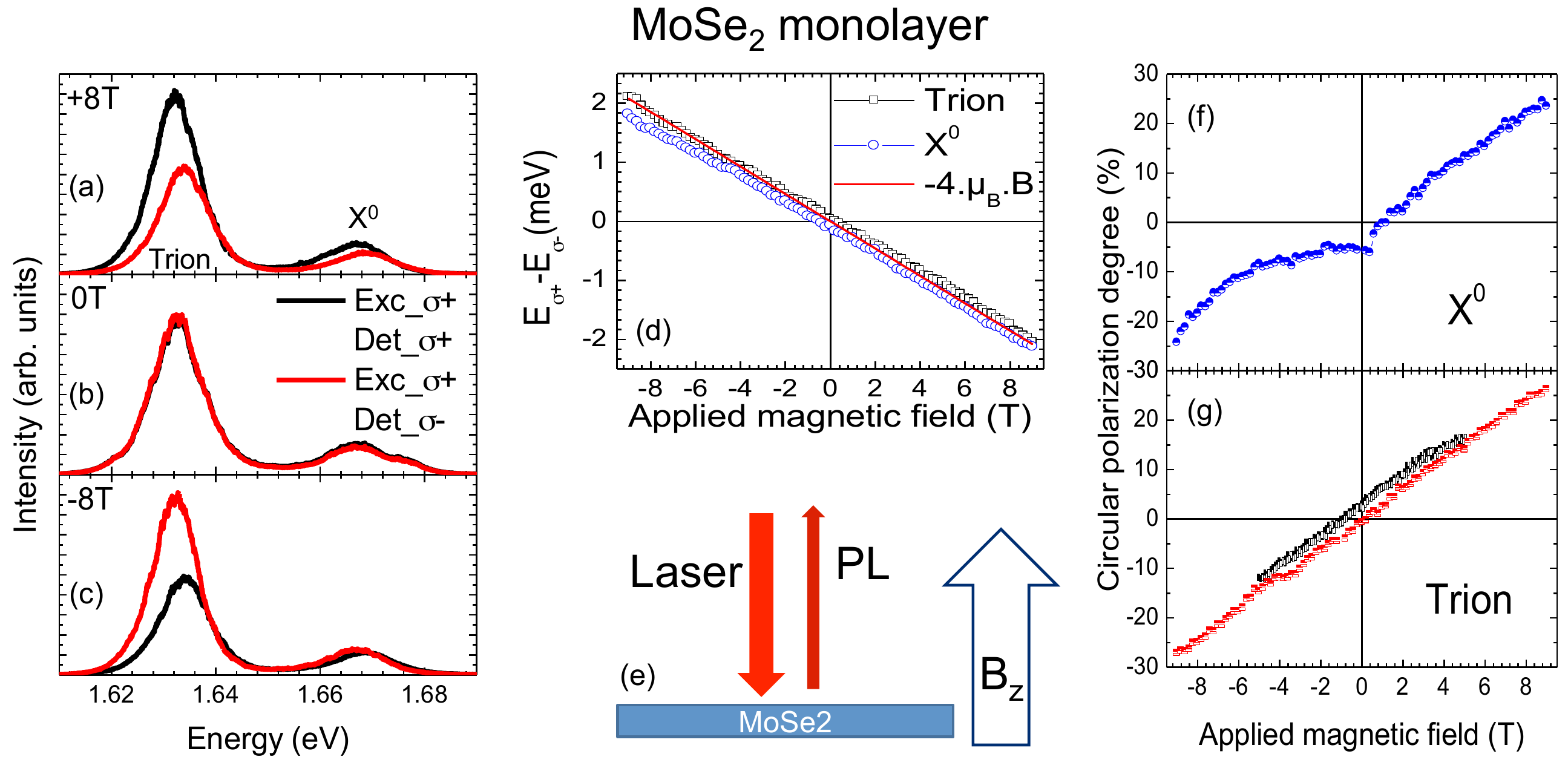}
\caption{\label{fig:fig1} \textbf{Data for monolayer MoSe$_2$}; excitation energy $E_\text{Laser}=1.96$~eV (a) PL spectra at $B_z=+8$~T for $\sigma^+$ (black) and $\sigma^-$ (red) polarized detection using $\sigma^+$ polarized laser excitation. (b) same as (a) but at $B_z=0$~T. (c) same as (a) but at $B_z=-8$~T. (d) Splitting between the $\sigma^+$ and $\sigma^-$ polarized PL components for the trion (black squares) and the X$^0$(blue circles) as a function of magnetic field, the function $-4\mu_B B_z$ is shown for comparison (red line). (e) Schematics of the experimental geometry. (f) Polarization of the PL emission of the X$^0$ as a function of $B_z$. (g) Polarization $P_c$ of the PL emission of the trion as a function of $B_z$ using $\sigma^+$  (black squares) and $\sigma^-$ excitation (red squares).
}
\end{figure*} 

\section{Experimental Set-up and Samples}
\label{sec:exp}

WSe$_2$ and MoSe$_2$ flakes  are obtained by exfoliation~\cite{Gomez:2014a} of a bulk crystal on a SiO$_2$/Si substrate. The ML regions are identified by optical contrast and very clearly in PL spectroscopy. Experiments at $T=4$~K and in magnetic fields up to $\pm$9~T have been carried out in an ultra-stable confocal microscope developed for spectroscopy on single semiconductor quantum dots (typical diameter of 20~nm) \cite{Bouet:2014a,Urbaszek:2013a}. We can infer that the mechanical movement of the detection and excitation spot on the sample due to the strong applied magnetic fields is at most in the tens of~nm range in our set-up. The detection spot diameter is about $700$~nm. The sample is excited either by a He-Ne laser (1.96~eV) or by a tunable continuous wave Ti:Sa laser.  The average laser power is in the $\mu$W range, in the linear absorption regime.  The PL emission is dispersed in a double-monochromator and detected with a Si-CCD camera.  The spectral resolution of this detection system is $\approx 20~\mu$eV. The circular PL polarization $P_c$ is defined as 
\begin{equation}
\label{circular}
{P_c=(I_{\sigma+}-I_{\sigma-})/(I_{\sigma+}+I_{\sigma-}),}
\end{equation}
 where $I_{\sigma+}(I_{\sigma-})$ denotes the intensity of the right ($\sigma^+$) and left ($\sigma^-$) circularly polarized emission. Light is $\sigma^+$ (right circularly polarized) if electric field vector rotates with time clock-wise provided one looks along the light propagation axis.  Similarly the linear PL polarization in the fixed $(xy)$ axes lying in the sample plane writes $P_l=(I_x-I_y)/(I_x+I_y)$ with $I_x(I_y)$ the $x$ and $y$ linearly polarized emission components.

\section{Magneto-optics in monolayer $\mbox{\textbf{MoSe}}_2$}
\label{sec:MoSe2}

First we discuss the experimental results for MoSe$_2$ MLs. At zero magnetic field, we observe in the PL spectrum two sharp emission features with a full width at half maximum (FWHM) of 10~meV. In accordance with previous reports \cite{Ross:2013a}, the low energy emission at 1.63~eV is attributed to the charged exciton (trion) recombination and at 1.67~eV we record the neutral exciton X$^0$ emission, see Fig.~\ref{fig:fig1}b. In our measurements the circular polarization degree $P_c$ of the trion and X$^0$ emission depends only very little on the excitation laser polarization ($\sigma^+$ or $\sigma^-$) for the excitation laser energy of 1.96~eV used in Fig.~\ref{fig:fig1}b. The maximum $P_c$ generated was of the order of 5\% for the excitation energy range investigated \cite{Wang:2015a}, much lower than the PL polarization achieved for circular laser excitation in MoS$_2$ \cite{Cao:2012a,Mak:2012a,Zeng:2012a,Sallen:2012a,Kioseoglou:2012a} and WSe$_2$ MLs \cite{Jones:2013a,Wang:2015b,Wang:2014b}. Below we discuss the magnetic field dependence of the PL emission, whose energy and polarization are essentially independent of the excitation laser polarization, in stark contrast to the results obtained in ML WSe$_2$ discussed in Sec. \ref{sec:WSe2}. 

Next we discuss the changes observed in the PL emission when applying a magnetic field $B_z$ perpendicular to the layer plane, i.e. along the $z$-direction, that is also the light propagation axis (Faraday geometry). Comparing Fig.~\ref{fig:fig1}a at $B_z=+8$~T and Fig.~\ref{fig:fig1}b at $B_z=0$ we observe two main differences: (i) For $B_z=+8~T$ the $\sigma^+$ polarized emission is more intense than the $\sigma^-$ polarized component, for both the trion and X$^0$; (ii) The  $\sigma^+$ polarized emission is shifted to lower energy compared to the $\sigma^-$ polarized component. When applying a field of $B_z=-8~$T in Fig.~\ref{fig:fig1}c, these results are reversed, i.e. $\sigma^-$ emission becomes more intense and lower in energy compared to $\sigma^+$ in agreement with time-reversal symmetry. In Fig.~\ref{fig:fig1}d we plot the full magnetic field dependence of the energy splitting 
\begin{equation}
\label{Zeeman}
{\Delta_Z=E^{PL}_{\sigma^+}-E^{PL}_{\sigma^-}=g \mu_B B_z,}
\end{equation}
 from $B_z=-9$~T to +9~T with $g$ being the effective $g$-factor and $\mu_B$ being the Bohr magneton.  The Zeeman splitting is extracted by fitting the trion and X$^0$ emission spectra with Lorentzians. We observe a clear linear dependence both for the trion and X$^0$ splitting on the applied field. This indicates that eventual diamagnetic and higher order contributions are identical for the upper and lower Zeeman branch within our experimental resolution. The slope is $\Delta_Z/B_z=-220\pm10~\mu$eV/T for the X$^0$ which corresponds to an exciton $g$-factor of $g_{X^0}=-3.8\pm0.2$. For the trion the slope is $-226\pm10~\mu$eV/T which corresponds to a $g$-factor  of $g_T=-3.9\pm0.2$. This corresponds to a maximum $\Delta_Z=-2$~meV at 9 Tesla. The main experimental uncertainty for $\Delta_Z$ comes from eventual changes of the overall shape of the PL emission due to imperfections in the optical set-up and sample inhomogeneities, as the recorded shifts are smaller than the linewidth. The Land\'{e}-factors for trions and the X$^0$ in ML MoSe$_2$ extracted from our date are close to the values reported in \cite{Macneill:2015a,Li:2014a}.\\
\indent It is worth mentioning that trions that form with the excess electron in the same or different valley, with respect to the photo-generated electron hole pair, are separated in energy due to the strong Coulomb effects and the zero-field splitting of conduction band states in each valley, cf.~\cite{Yu:2014b}. However, we do not observe any fine structure splitting for the trion emission in the investigated sample. Hence, the measured $g_T$ represents the global magnetic field induced energy shift of the trion emission. \\
\indent We now discuss the circular polarization of the observed emission summarized in Figs.~\ref{fig:fig1}f and \ref{fig:fig1}g. For the trion, $P_c$ increases from zero to 30\% when the magnetic field is increased from zero up to 9~T. This strong polarization is reversed when the direction of the applied magnetic field is reversed. Optical valley initialization at higher fields is not at the origin of this effect: For $\sigma^+$ polarized laser excitation we observed exactly the same polarization increase with field as for $\sigma^-$ polarized laser excitation. We conclude that the PL polarization is the result of magneto-induced spin/valley polarization build-up during the PL emission time, which has been determined to be in the ps-range \cite{Wang:2015a}. Observing an increase in the emission polarization in applied magnetic fields is a very common observation in semiconductors \cite{Dyakonov:2008a}. For ML MoSe$_2$, where excitons have a strong binding energy \cite{Ugeda:2014a}, this observation is very surprising taking into account the relevant energy scales. At 9 Tesla, we have induced a splitting between the valley Zeeman levels of 2~meV, eventually comparable to the spin splitting in the conduction band, predicted to be in the meV range \cite{Kosmider:2013a,Liu:2013a}. The polarization of the neutral exciton follows a similar trend as the trion. For the neutral exciton, the applied field $B_z$ can dominate the long-range electron-hole Coulomb exchange interaction, and hence suppress valley depolarization when $B_z$ has a larger amplitude than the effective transverse field associated to exchange effects \cite{Maialle:1993a,Glazov:2014a,Yu:2014a,Yu:2014b}. 

\section{Magneto-optics in monolayer $\mbox{\textbf{WSe}}_2$}
\label{sec:WSe2}

 \begin{figure*}[t]
\includegraphics[width=0.95\textwidth]{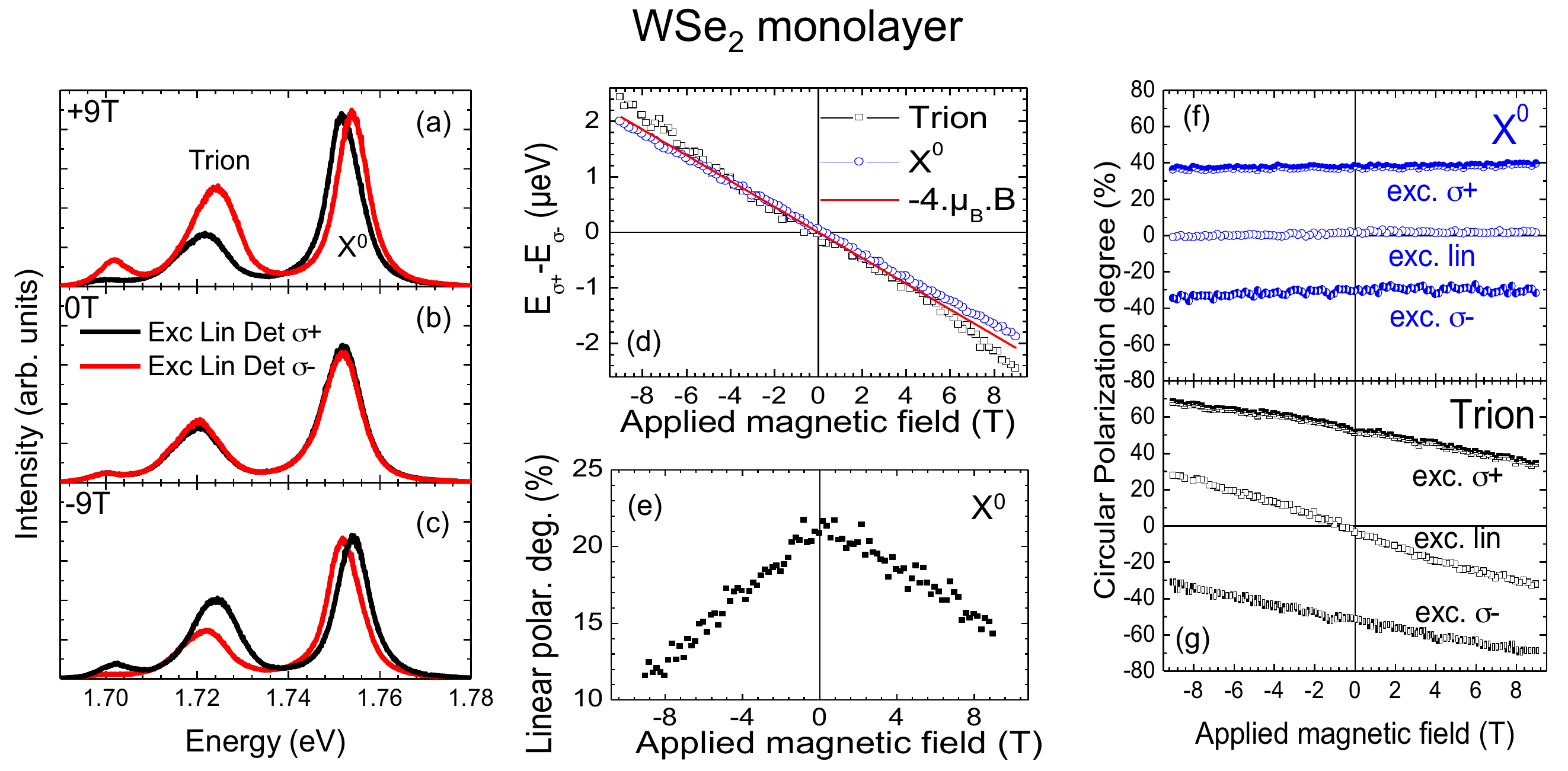}
\caption{\label{fig:fig2} \textbf{Data for monolayer WSe$_2$}; excitation energy $E_\text{Laser}=1.96$~eV (a) PL spectra at $B_z=+9$~T for $\sigma^+$ (black) and $\sigma^-$ (red) polarized detection using $\sigma^+$ polarized laser excitation. (b) same as (a) but at $B_z=0$~T. (c) same as (a) but at $B_z=-9$~T. (d) Splitting between the $\sigma^+$ and $\sigma^-$ polarized PL components for the trion (black squares) and the X$^0$(blue circles) as a function of magnetic field, the function $-4\mu_B B_z$ is shown for comparison (red line). (e) Linear polarization of the X$^0$ emission as a function of $B_z$. (f) Polarization of the PL emission of the X$^0$ as a function of $B_z$ for three different laser polarizations. (g) Polarization $P_c$ of the PL emission of the trion as a function of $B_z$ for three different laser polarizations.
}
\end{figure*}

In order to draw more general conclusions on valley properties in applied magnetic fields, we compare our results in ML MoSe$_2$ with a very well characterized system, namely ML WSe$_2$ \cite{Jones:2013a,Wang:2015b,Wang:2014b} investigated with the same experimental set-up. Therefore all experimental uncertainties are comparable. In Fig.~\ref{fig:fig2}b the PL emission of ML WSe$_2$ is plotted, and just as in the case of ML MoSe$_2$, we observe two well defined emission lines with a typical FWHM of 10~meV. The higher energy transition at 1.75~eV is identified as the neutral exciton X$^0$ recombination, as valley coherence can be generated \cite{Jones:2013a}, see Fig.~\ref{fig:fig2}e. 
The lower energy transition at 1.72~eV stems from the radiative recombination of the charged exciton (trion).  The $\sigma^+$ and $\sigma^-$ polarized PL components correspond to carrier recombination in the $K^+$ and $K^-$ valley, respectively. At zero magnetic field the emission spectra in $\sigma^+$ and $\sigma^-$ polarizations are exactly the same due to the time-reversal symmetry and corresponding degeneracy of the valley levels.
 In a magnetic field of 9~T this valley degeneracy is lifted, and the energy difference $\Delta_Z=E^{PL}_{\sigma^+}-E^{PL}_{\sigma^-}$, Eq.~\eqref{Zeeman}, is typically $-2$~meV, see Fig.~\ref{fig:fig2}a, very similar to the results on MoSe$_2$ presented in Fig.~\ref{fig:fig1}a. In a field of $B_z=-9$~T, the $\sigma^+$ polarized component is now at higher energy, corresponding to a positive Zeeman splitting of +2~meV. 
 \begin{table}
\caption{\label{tab:table1} comparison of Land\'e factors }
\begin{ruledtabular}
\begin{tabular}{ccccc}  
  &$g_{X^0}$ & X$^0$ slope & $g_T$ & Trion slope   \\
  & &[$\mu$eV/T]& &[$\mu$eV/T] \\
\hline
ML MoSe$_2$ & $-3.8\pm0.2$ &	$-220\pm10$ &	$-3.9\pm0.2$ & $-226\pm10$  \\
ML WSe$_2$ & $-3.7\pm0.2$ &	$-214\pm10$	& $-4.4\pm0.2$ & $-254\pm10$  \\
\end{tabular}
\end{ruledtabular}
\end{table} 
We plot in Fig.~\ref{fig:fig2}d the full magnetic field dependence of the Zeeman splitting for the trion and X$^0$. The experimental points are very close to a straight line, which allows us to extract for the X$^0$ a slope of $\Delta_Z/B_z=-214\pm10~\mu$eV/T, corresponding to the neutral exciton $g$-factor $g_{X^0}=-3.7\pm0.2$. For the trion the slope is $-254\pm10~\mu$eV/T, resulting in its $g$-factor $g_T=-4.4\pm0.2$.
  
The similarities between the measured Land\'e-factors in MoSe$_2$ and WSe$_2$ are striking, both with respect to their signs and amplitudes, see table \ref{tab:table1}. Comparing with the very recent literature data on ML WSe$_2$, the $g$-factors for the neutral exciton extracted for our sample are smaller than reported in reference \cite{Srivastava:2015a} and larger than in reference \cite{Aivazian:2015a}. Note, that as for MoSe$_2$ case we do not resolve any fine-structure of the trion emission in WSe$_2$.

Although the Zeeman energy evolution of the both ML materials MoSe$_2$ and WSe$_2$ shows close similarities, we will see below that the evolution of the valley polarization as a function of the applied magnetic field is completely different. The polarization measurements on WSe$_2$ are all performed with non-resonant HeNe laser excitation. First, we plot the evolution of the X$^0$ PL polarization under polarized pumping, that corresponds to the valley polarization generation via the chiral optical selection rules \cite{Xiao:2012a,Cao:2012a}, as a function of the applied magnetic field. It can be seen in Fig.~\ref{fig:fig2}f that the X$^0$ polarization is essentially independent of the applied field. For a linearly polarized excitation laser, we find $P_c \approx 0$ for all applied field values. Changing to $\sigma^+$ circularly polarized excitation, we record at zero Tesla $P_c\approx 40\%$. This value remains practically constant in applied fields from $-9$~T to $+9$~T. This indicates that over the entire magnetic field range, the X$^0$ emission polarization is determined by the initially, optically created valley polarization rather than induced by the magnetic field. During the X$^0$ PL emission time no valley/spin relaxation occurs, which might simply be a consequence of the extremely short PL emission time in the few ps range measured for this material \cite{Wang:2014b}. Exciting preferentially the $K^-$ valley with a $\sigma^-$ polarized laser, results in $P_c\approx-40\%$, again independent of the applied magnetic field.  This behaviour for the X$^0$ in ML WSe$_2$ is in stark contrast to the observations in ML MoSe$_2$. For the former, the optical valley initialization determines the PL polarization, for the latter the applied magnetic field direction and amplitude allow to control the valley polarization. Comparing with the recent literature, the X$^0$ polarization in the ML WSe$_2$ sample investigated by \textcite{Aivazian:2015a}  was slightly more sensitive to the applied magnetic field. 

Contrary to the X$^0$ in our ML WSe$_2$ sample, the trion polarization in this ML can be controlled via the applied magnetic field, shown in Fig.~\ref{fig:fig2}g. For linearly polarized laser excitation, the trion PL polarization for $B_z=0$ is absent, $P_c=0$, but increases at $-9$~T to $+30\%$, and at $+9$~T to $-30$\%. This can be directly compared to the trion in ML MoSe$_2$, that also started off at zero field with zero polarization. When comparing the trion polarization evolution in ML MoSe$_2$ (Fig.~\ref{fig:fig1}g) and ML WSe$_2$ (Fig.~\ref{fig:fig2}g) we note very contrasting behaviour: Application of a positive magnetic field, results in a strong, positive $P_c$ in MoSe$_2$ in contrast to the strong, but negative polarization created in WSe$_2$. Using $\sigma^+$ polarized excitation for the trion, results in high circular polarization $P_c=50\%$ already at $B_z=0$. At a magnetic field of $B_z=-9$~T this polarization is increased to $P_c=68\%$, at $B_z=+9$~T we find $P_c=34\%$. For the trion using $\sigma^+$ excitation, we find a similar trend as under linearly polarized excitation: the PL polarization decreases for positive $B_z$ and increases for negative $B_z$. The trion polarization depends on both the excitation laser polarization and the applied magnetic field. This latter dependence might be linked to the fact that the PL emission time of the trion in ML WSe$_2$ is longer than for the X$^0$ \cite{Wang:2014b}, allowing for polarization relaxation to occur before radiative recombination takes place. As can be seen in Fig.~\ref{fig:fig2}g, the magnetic field induced change in PL polarization is also observed for $\sigma^-$ excitation, here we find at $B_z=-9$~T $P_c=-31\%$, while for zero field we record $P_c=-52\%$ (as expected when switching from $\sigma^+$ to $\sigma^-$ excitation) evolving towards $P_c=-69\%$ for $B_z=-9$~T. The three different measurement series presented in Fig.~\ref{fig:fig2}g each confirm that the PL polarization of the trion in ML WSe$_2$ strongly \textit{decreases} with the applied magnetic field, whereas, in stark contrast, the trion PL polarization in ML MoSe$_2$ strongly \textit{increases} as a function of the applied $B_z$.

\section{Discussion}
\label{sec:dis}

The emission energies and polarizations are determined by several factors. Most importantly, the Zeeman splittings of exciton and trion in the emission spectra are governed by an interplay of spin splittings of conduction and valence band states, while the polarization is governed by (i) selection rules at optical transitions, (ii) occupancies of the spin/valley states and possible spin relaxation processes. Here we briefly discuss theoretical approaches to evaluate Zeeman splittings and polarization of emitted radiation in the context of experimental data presented above.

\subsection{Zeeman effect in two-dimensional crystals}
\label{sec:Zeeman}

We recall that the electron wavefunction in the two-dimensional crystal can be recast, in accordance with the Bloch theorem, as
\begin{equation}
\label{Bloch}
\psi_{n\bm q}(\bm r) = \frac{\mathrm e^{{\rm i} {\bm q} {\bm r}}}{\sqrt{S}} u_{n{\bm q}}({\bm r}),
\end{equation}
where $n$ enumerates bands (including electron spin state) and $\bm q$ is the quasi-wavevector, $u_{n\bm q}(\bm r)$ is the periodic amplitude normalized per volume of the unit cell, $\Omega_0$: $\int_{\Omega_0}|u_{n\bm q}|^2 d\bm r = \Omega_0$, and $S$ is the macroscopic normalization area.

The bare electron Zeeman effect is described by
\begin{equation}
\label{bare:Z}
\mathcal H_0 = g_0 \mu_B \bm B \cdot \bm s,
\end{equation}
where $\mu_B=|e|\hbar/(2m_0c)$ is the Bohr magneton, $m_0$ and $g_0=2$ are the free electron mass and Land\'e factor, and $\bm s$ is the spin operator $\bm \sigma/2$, $\bm \sigma$ being the vector composed of the Pauli matrices. In crystals, in addition to (\ref{bare:Z}) the orbital contribution to the Zeeman splitting should be taken into account. It is related to the orbital momentum of the electron~\cite{ll3_eng},
\begin{equation}
\label{orb:Z}
\mathcal H_1 = \mu_B \bm B \cdot \bm L,
\end{equation}
where $\bm L = \hbar^{-1}[\bm r \times \bm p]$ is the angular momentum operator, and $\bm p = -\mathrm i \hbar\bm \nabla$ is the electron momentum operator. The contribution~\eqref{orb:Z} is known to be important for well characterized semiconductors such as GaAs~\cite{Ivchenko:1995a,ivchenko05a}. 

We recall that in MoSe$_2$ and WSe$_2$ MLs the direct band gaps are realized at the edges of the Brillouin zone characterized by the wavevectors $\bm K^\pm$. The bands at $\bm q = \bm K^{\pm}$ are non-degenerate and can be characterized by a certain spin projection $s_z = \pm 1/2$ onto the sample normal. The time reversal symmetry couples $s_z = \pm 1/2$ states in the $\bm K^+$ valley with $s_z = \mp 1/2$ states in the $\bm K^-$ valley. In what follows we will be interested in the Zeeman effect in magnetic field $\bm B \parallel z$. Correspondingly, it is instructive to present the diagonal matrix element of $L_z$ operator at $\bm q = \bm K^{+}$ or $\bm K^-$ as
\begin{multline}
\label{Lz}
\langle \psi_n|\hbar L_z|\psi_n\rangle_\pm = \\
\sum_{m\ne n}[\Omega_{nm}^x(\bm K^{\pm})p_{mn}^y(\bm K^{\pm}) - \Omega_{nm}^y(\bm K^{\pm})p_{mn}^x(\bm K^{\pm}) ].
\end{multline} 
Here $p^\alpha_{nm}$ ($\alpha=x,y$) are the  matrix elements of the electron momentum and $\Omega_{nm}^\alpha = \mathrm i \Omega_0^{-1} \int_{\Omega_0} u_n^* \left( \partial u_m/\partial q_\alpha\right)  d\bm r$ are the interband matrix elements of the coordinate operator. In derivation of Eq.~\eqref{Lz} we made use of the completeness relation for the Bloch amplitudes and took into account that $p_{nn}(\bm K^{\pm})=0$. Taking into account that $\Omega_{mn}^\alpha = \mathrm i \hbar p_{mn}^\alpha/[(E_n - E_m)m_0]$ with $E_n$, $E_m$ being the energies of corresponding bands, Eq.~\eqref{Lz} can be rewritten as~\cite{ivchenko05a}
\begin{multline}
\label{Lz1}
\langle \psi_n| L_z|\psi_n\rangle_\pm = \\
\frac{\mathrm i}{m_0}\sum_{m\ne n}\frac{p_{nm}^x(\bm K^{\pm})p_{mn}^y(\bm K^{\pm}) - p_{nm}^y(\bm K^{\pm})p_{mn}^x(\bm K^{\pm})}{E_m - E_n}.
\end{multline} 
Equations~\eqref{bare:Z}, \eqref{orb:Z} and \eqref{Lz1} can be used to evaluate Zeeman splittings both in the $\bm k\cdot \bm p$ and tight-binding models, as detailed in the next two subsections.

\subsection{kp-theory}\label{sec:kp}

The multiband $\bm k \cdot \bm p$ model was formulated for TMDCs in Refs.~\cite{Kormanyos:2013a,Kormanyos:2014a,Kormanyos:2014b}. The effective Hamiltonians describing the states in the vicinity of $\bm K^\pm$ edges of the Brillouin zone have the form
\begin{subequations}
\label{Hkp}
\begin{equation}
\label{Hkp+}
\mathcal H_+ =\begin{pmatrix}
E_{c+2} & \gamma_6 k_- & \gamma_4 k_+ & 0 \\
\gamma_6 k_+ & E_c & \gamma_3 k_- & \gamma_5 k_+ \\
\gamma_4 k_- & \gamma_3 k_+ & E_v & \gamma_2 k_- \\
0 & \gamma_5 k_- & \gamma_2 k_+ & E_{v-3}
\end{pmatrix}, 
\end{equation}
\begin{equation}
\label{Hkp-}
\mathcal H_- =  \begin{pmatrix}
E_{c+2} & \gamma_6 k_+ & \gamma_4 k_- & 0 \\
\gamma_6 k_- & E_c & \gamma_3 k_+ & \gamma_5 k_- \\
\gamma_4 k_+ & \gamma_3 k_- & E_v & \gamma_2 k_+ \\
0 & \gamma_5 k_+ & \gamma_2 k_- & E_{v-3}
\end{pmatrix}.
\end{equation}
\end{subequations}
Here $k_\pm = k_x \pm \mathrm i k_y$ are the cyclic components of the electron wavevector reckoned from the $\bm K^\pm$ points, $\bm k_{\pm} = \bm K^\pm - \bm q$; the parameters $\gamma_3 \ldots \gamma_6$ are related to the interband momentum matrix elements, and certain convention about the phases of the Bloch functions is assumed. The symbols $c+2$ and $v-3$ denote excited conduction and deep valence bands. Such a model was shown to allow for adaquate description of the spin-orbit coupling and trigonal symmetry effects in TMDCs~\cite{Kormanyos:2013a,Kormanyos:2014a,Kormanyos:2014b}.
In the general theory besides $\bm k$-linear off-diagonal terms, effective $\bm k\cdot \bm p$ Hamiltonian includes diagonal quadratic in $\bm k$ contributions resulting (i) from the second-order $\bm k\cdot \bm p$ coupling with distant bands and (ii) from bare electron dispersion $\hbar^2k^2/2m_0$~\cite{birpikus_eng}. Usually these contributions are comparable and should be included simultaneously. Diagonalizing Hamiltonians~\eqref{Hkp} in the second order in off-diagonal terms we obtain for the electron effective masses of the conduction and valence bands, respectively,
\begin{subequations}
\label{mass:kp1}
\begin{eqnarray}
\frac{1}{m_c} &=& \frac{1}{m^*} + \frac{1}{m_c'}+ \frac{1}{m_0} +\frac{1}{m_c''},\\
\frac{1}{m_v} &=& -\frac{1}{m^*} + \frac{1}{m_v'}+ \frac{1}{m_0} + \frac{1}{m_v''}.
\end{eqnarray}
\end{subequations}
Here subscripts $c, v$ denote the corresponding bands and we use the electron representation, $m^* = \hbar^2 (E_c - E_v)/(2\gamma_3^2)$ is the two-band effective mass, the terms
\begin{subequations}
\label{5band}
\begin{equation}
\frac{1}{m_c'} = \frac{2}{\hbar^2}\left(\frac{\gamma_5^2}{E_c - E_{v-3}} + \frac{\gamma_6^2}{E_c - E_{c+2}}\right),
\end{equation}
\begin{equation}
\frac{1}{m_v'} = \frac{2}{\hbar^2}\left(\frac{\gamma_2^2}{E_v - E_{v-3}} + \frac{\gamma_4^2}{E_v - E_{c+2}}\right)
\end{equation}
\end{subequations}
result from the mixing described by the Hamiltonians~\eqref{Hkp} and $1/m_c''$, $1/m_v''$ contain above mentioned contributions from remote bands [not included in Eqs.~\eqref{Hkp}].

According to the general theory~\cite{birpikus_eng} the magnetic field within the $\bm k \cdot \bm p$ scheme is included (i) by adding the bare Zeeman effect in the form of Eq.~\eqref{bare:Z} and (ii) by replacing $\bm k$ in Eqs.~\eqref{Hkp} by $\bm k - (e/c\hbar) \bm A$, where $e=-|e|$ is the electron charge,  $\bm A$ is the vector potential of the magnetic field. The calculation in the first order in $B_z$ yields the effective $g$-factors of electrons in $\bm K^\pm$ valleys:
\begin{equation}
\label{g+}
g_{c,v}^{\bm K^+} \equiv g_{c,v} =  {2 + g^{\rm orb}_{c,v} }, \quad g_{c,v}^{\bm K^-} = -g_{c,v}^{\bm K^+}.
\end{equation}
The term $2$ in Eq.~\eqref{g+} arises from the bare Zeeman effect, Eq.~\eqref{bare:Z}, while $g_{c,v}^{orb}$ result from the $\bm k \cdot \bm p$-mixing
\begin{subequations}
\label{g:orb}
\begin{equation}
\label{g:kp:c}
g_c^{\rm orb} = \frac{4m_0}{\hbar^2}\left(\frac{\gamma_3^2}{E_c - E_v} - \frac{\gamma_5^2}{E_c - E_{v-3}} - \frac{\gamma_6^2}{E_c - E_{c+2}}\right),
\end{equation}
\begin{equation}
\label{g:kp:v}
g_v^{\rm orb} = \frac{4m_0}{\hbar^2}\left(-\frac{\gamma_3^2}{E_v - E_c} + \frac{\gamma_2^2}{E_v - E_{v-3}} + \frac{\gamma_4^2}{E_v - E_{c+2}}\right).
\end{equation}
\end{subequations}
Equations~\eqref{g:orb} can be derived from Eqs.~\eqref{orb:Z} and \eqref{Lz1}, taking into account only bands $c$, $c+2$, $v$ and $v-3$. Additional contribution to $g^{orb}_{c,v}$, namely, $\Delta g^{orb}_{c,v}$ may arise allowing for other distant bands in Eq.~\eqref{Lz1}.

\subsection{Tight-binding theory}\label{sec:tb}

In the tight-binding approximation the Bloch function in Eq.~\eqref{Bloch} is presented as a linear combination of atomic orbitals $\phi_j^a(\bm r )$ (generally orthogonalized~\cite{lowdin}) in the form
\begin{equation}
\label{TB}
\psi_{n\bm q}(\bm r) = \sum_{a,l,j} \mathrm e^{\mathrm i \bm q \bm R_{a,l}} C_{l,j}^a \phi_j^a(\bm r - \bm R_{a,l}).
\end{equation}
Here $a$ enumerates types of atoms ($a=$ Metal or Chalcogen), $l$ enumerates atoms of a given type, $j$ runs through the set of orbitals taken into account at a given atom, $\bm R_{a,l}$ are the positions of the atoms in the two-dimensional lattice and $C_{l,j}^a \equiv C_{l,j}^a (n, \bm q)$ are the coefficients. The tight-binding Hamiltonian acting in the space of coefficients $C_{j,l}^a$ contains diagonal energies, i.e. energies of orbitals, and hopping matrix elements mixing coefficients $C_{l,j}^a$ and $C_{l',j'}^{a'}$ for different atoms ($l \ne l'$). There are several tight-binding models developed for TMDC monolayers~\cite{C4CS00301B}: (i) the $3$-band model~\cite{Liu:2013a} which takes into account only three $d$-orbitals of Metal, namely, $d_{z^2}$, $d_{x^2-y^2}$, and $d_{xy}$, and includes up to three nearest neighbours to reproduce density functional theory (DFT) calculations of electron/hole dispersion over the whole Brillouin zone, (ii) the $7$-band model~\cite{Rostami:2013oq} which includes three $d$-orbitals of Metal, two $p$-orbitals for each Chalcogen and only nearest neighbour interactions (this model involves as well overlap matrix elements since it is developed for non-orthogonalized orbitals); (iii) the $11$-band model of Ref.~\cite{PhysRevB.88.075409} which accounts for five $d$-orbitals of Metal and three $p$-orbitals of each Chalcogen and applies the nearest neighbour approximation; (iv) the $27$-band model of Ref.~\cite{:/content/aip/journal/adva/3/5/10.1063/1.4804936} which takes into account $sp^3d^5$ states for each atom and provides rather high accuracy.

The diagonalization of the tight-binding Hamiltonian provides the energy dispersion in the whole Brillouin zone. In the vicinity of $\bm K^\pm$ points the tight-binding Hamiltonian can be expanded in power series in the electron wavevector (referred to the Brillouin zone edge) and can be partially diagonalized. The resulting effective Hamiltonian has a form similar to the $\bm k\cdot \bm p$ Hamiltonian (with different number of bands depending on the tight-binding model) but contains diagonal $k^2$ contributions. It is convenient to interpret the latter as an effective 
$\bm k\cdot \bm p$ contribution of ``the bare electron mass and remote bands''. The off-diagonal elements (linear in $\bm k$) contain coefficients which can be interpreted as inter-center contributions to the momentum matrix elements~\cite{PhysRevB.47.15500,PhysRevB.51.4940,goupalov2001tight,Xu19901143}.
For example, the $3$-band model of Ref.~\cite{Liu:2013a} yields a $3\times 3$ effective Hamiltonian which includes the $c$, $v$ and $c+2$ bands in notations of the $\bm k\cdot \bm p$ Hamiltonian~\eqref{Hkp}. The $7$-band model of Ref.~\cite{Rostami:2013oq} yields an additional conduction band with the same symmetry as the $c$-band.

Within the tight-binding approach the magnetic field is also included in a two-fold way. First, the phase of the hopping matrix elements is modified by including the vector potential as follows
\begin{equation}
\label{phase}
-\frac{\mathrm i e}{c\hbar}\int_{\bm r_1}^{\bm r_2} \bm A(\bm r) \mathrm d\bm r.
\end{equation} 
This procedure is equivalent to the $\bm k \to \bm k - (e/c\hbar) \bm A$ replacement in the $\bm k \cdot \bm p$ Hamiltonian. Second, the intra-atomic contribution should be included. It contains the spin part, Eq.~\eqref{bare:Z}, as well as possible orbital contribution. The latter should be carefully calculated using  the restricted basis of atomic orbitals, the completeness relation for this restricted basis, and an analogue of Eq.~\eqref{Lz} in the form 
\begin{eqnarray}
\label{completeness}
&& \langle \phi_j^a(\bm r)| \hbar L_z |\phi_j^a(\bm r)\rangle  \\
&&= \sum_{i} ( \langle \phi_j^a|\hat x|\phi_{i}^a \rangle \langle \phi_{i}^a| \hat p_y |\phi_j^a\rangle - \langle \phi_j^a|\hat y|\phi_{i}^a \rangle \langle \phi_{i}^a | \hat p_x |\phi_j^a\rangle)\:.
\nonumber
 \end{eqnarray}
Note, that the summation in Eq.~\eqref{completeness} extends over the orbitals used in Eq.~\eqref{TB} rather than over all possible states of an isolated atom, otherwise the problem becomes ill-defined. 
 
In the particular cases of $3$-, $7$- and $11$-band tight-binding models, the intra-site values of the momentum operator vanish due to symmetry reasons because for any given atom either only $p$-shell or $d$-shell states are included into the model. Therefore, in Eq.~\eqref{completeness} $\langle \phi_j^a(\bm r)| \hbar L_z |\phi_j^a(\bm r)\rangle=0$ and the intra-site contribution to the Zeeman effect vanishes. In this case we obtain within the tight-binding method Eq.~\eqref{Lz1} for the orbital momentum and recover expressions analogous to Eq.~\eqref{g+}, Eqs.~\eqref{g:orb} from $\bm k\cdot\bm p$ theory with no additional intra-center contributions in contrast to Refs.~\cite{Macneill:2015a,Li:2014a,Aivazian:2015a}, see also Ref.~\cite{note:GaAs}. Additional intra-site contributions may arise in the advanced tight-binding formalism of Ref.~\cite{:/content/aip/journal/adva/3/5/10.1063/1.4804936} where orbitals of different symmetry are included for a given atom, in which case the momentum operator matrix elements between these orbitals may become non-zero.

\subsection{Zeeman splittings of direct excitons and trions}

The neutral exciton radiative decay involving emission of $\sigma^+$ or $\sigma^-$ photons results from the recombination of a Coulomb-correlated electron-hole pair in the $\bm K^+$ or $\bm K^-$ valley, respectively. The Zeeman splitting of X$^0$ is, in accordance with Eq.~\eqref{Zeeman}, given by
\begin{equation}
\label{splittings:X0}
\Delta_Z = \frac{1}{2} \left[ g_c^{\bm K_+}  - g_v^{\bm K_+} - \left( g_c^{\bm K_-} -  g_v^{\bm K_-} \right)\right] \mu_B B_z\:.\end{equation}
Note that hereafter we neglect the renormalization of $g$-factor due to the Coulomb effects and band non-parabolicity. Making use of Eqs.~\eqref{g+}, \eqref{g:orb} the bright exciton $g$-factor is given by
\begin{equation}
\label{g:exc}
g_x = g_c - g_v = -2\left(\frac{m_0}{m_c'} +\frac{m_0}{m_v'}\right) + \Delta g_c^{\rm orb} -\Delta g_v^{\rm orb},
\end{equation}
where terms in parenthesis are calculated within the framework of Hamiltonian \eqref{bare:Z} and 4-band Hamitonian~\eqref{Hkp}, and the contribution $\Delta g_c^{orb} -\Delta g_v^{orb}$ results form the remote bands not accounted for by Eq.~\eqref{Hkp}.  
At this stage precise measurements of the conduction band effective masses have not been reported yet. First measured values for the valence band effective masses can be obtained from angle-resolved photoemission spectroscopy (ARPES) \cite{Zhang:2014a,Riley:2014a}.
Equation~\eqref{g:exc} can also be represented in the form 
\begin{eqnarray}
\label{g:exc1}
g_x &=&  4 - 2 \left( \frac{m_0}{m_c} +\frac{m_0}{m_v} - \frac{m_0}{m_c''} - \frac{m_0}{m_v''} \right) \nonumber\\ &&~~+~ \Delta g_c^{\rm orb} -\Delta g_v^{\rm orb}\:,
\end{eqnarray}
where the contributions to the Land\'{e} factor and effective masses resulting from remote bands are explicitly present. 
Note that in the two-band approximation the bright exciton $g$-factor exactly vanishes because in this approximation the conduction- and valence-band electron effective masses are given by $m_0/m_c^{2b} = 1+ m_0/m^*$, $m_0/m_v^{2b} = 1-m_0/m^*$ and, therefore, 
\[
g^{2b}_x = 4 - 2\left(\frac{m_0}{m_c^{2b}} +\frac{m_0}{m_v^{2b}}\right) = 4-2\times 2=0.
\] 
The term 4 arises from inclusion of $1/m_0$ in the inverse effective masses for the conduction and valence band electrons.  It should be emphasized that the inclusion of free electron dispersion $1/m_0$ in Eqs.~\eqref{mass:kp1} and neglecting the terms $1/m_c''$, $1/m_v''$, $\Delta g_c^{orb}$, and $\Delta g_v^{orb}$ contributed by remote bands cannot formally be justified within the $\bm k \cdot \bm p$ method. Those contributions should be estimated from experimental data or evaluated via more advanced $\bm k\cdot \bm p$ schemes or atomistic approaches. 

Since in MoSe$_2$ and WSe$_2$ the conduction- and valence-band effective masses are close in absolute values but have opposite signs, in accordance with Eqs.~\eqref{g:exc}, \eqref{g:exc1}, the orbital contribution to the exciton $g$-factor stems from remote bands. We note that neither Eq.~\eqref{g:exc} with $\Delta g_c^{\rm orb} -\Delta g_v^{\rm orb}=0$ nor Eq.~\eqref{g:exc1} with $m_v'', m_c''\to \infty$ can satisfactory describe the experiment in respect of both the sign and the magnitude of $g$-factor. Hence we conclude that the remote band contributions are important. For instance, if we add into consideration one more (distant) band with the same symmetry as the conduction band with band edge energy $E_c'$ and the matrix element of $\bm k\cdot\bm p$ interaction with the valence band $\gamma_3'$ we obtain an additional contribution to the valence band $g$-factor $\Delta g_v = 4m_0\gamma_3'^2/[\hbar^2(E_c' - E_v)]$, while $\Delta g_c =0$ for the symmetry reasons. Further experimental studies and theoretical modelling are therefore needed to elucidate the values of Zeeman splitting.

The Zeeman splitting for the bright trion is also given by Eq.~\eqref{splittings:X0}. This is because the optical recombination involves charge carriers of opposite signs in the same valley, while the spin/valley state of the third carrier is not changed. The difference of measured values for excitons and trions can be attributed to the Coulomb-induced renormalization of $g$-factors due to band non-parabolicity.

\subsection{Polarization of emission}\label{sec:pol}

The polarization of the PL emission is governed by the selection rules and occupancies of spin/valley states of the carriers and Coulomb complexes. In case of MoSe$_2$, neither optical orientation (circular polarization of photoluminescence for circularly polarized excitation at $\bm B=0$) nor exciton alignment or valley coherence (linear polarization at linearly polarized excitation) are observed \cite{Wang:2015a}. This allows us to assume that in this material the spin/valley relaxation of both neutral excitons and trions (or individual carriers) is fast compared to the PL emission time. In the presence of a magnetic field the magneto-induced circular polarization seems to result from the preferential occupation of the lowest Zeeman state of the exciton or trion. It is indeed consistent with experiment where, for $B_z>0$, the state emitting the $\sigma^+$-polarized photons has the lower energy and is dominant in the photoluminescence so that $P_c(B_z>0)>0$, see Fig.~\ref{fig:fig1}. Note, however, that the full thermalization to the lattice temperature does not occur because at $|B_z|\approx 9$~T the Zeeman splitting exceeds by far the temperature expressed in the energy units, while experiment demonstrates only $|P_c|\sim 30$~\%. The effective spin temperature deduced from the experiment is about $T_\text{spin}=30$~K.

The situation is more complex for WSe$_2$ where the experiment shows a substantial optical orientation, Fig.~\ref{fig:fig2}f, and neutral-exciton alignment, Fig.~\ref{fig:fig2}e, even at $B_z=0$.  In this case the spin/valley relaxation time is comparable to the lifetime of excitations and, in an applied magnetic field, thermalization may not occur. Additionally, as compared to MoSe$_2$, WSe$_2$ is characterized by the opposite sign of zero-field spin splitting of the conduction band~\cite{Kosmider:2013a,Liu:2013a}. Hence, for the exciton and trion ground states the direct intra-valley optical transitions are spin-forbidden. These states can be manifested in optical spectra due to indirect (e.g. phonon-assisted) transitions similarly to the case of Carbon nanotubes~\cite{PhysRevLett.101.087402}. It follows from the symmetry considerations that in this case the polarization is reversed ~\cite{note:gi:indirect}. Further experimental data using gated devices where positively and negatively charged excitons can be clearly distinguished and theoretical analysis are needed to clarify this issue.

Finally we emphasize that the above analysis is based on the perturbative treatment of the Coulomb interaction while, in TMDC MLs, the exciton binding energy amounts to $\gtrsim 0.5$~eV and, hence, is comparable to the band gap. Theoretical estimates show that depending on the parameters of the materials and dielectric environment the binding energy is so large that the $1s$ state emission could be in the infra-red range and, moreover, the exciton ground state could even collapse~\cite{Rodin:2013a,Stroucken:2014a}. In the former case the optical transition $2p\to 1s$ could be relevant as well, its Zeeman splitting and polarization deserve further study. In the latter case the ground state of the system could be strongly renormalized and an excitonic insulator could be formed~\cite{keldysh1965possible,KozlovMaksimov,PhysRev.158.462,GuseinovKeldysh}. Its polarization and magnetic field properties should also be studied in future works.

\indent \textit{Acknowledgements.---} We thank Andor Kormanyos, Hanan Dery, Iann Gerber and Junichiro Kono for very fruitful discussions. We acknowledge partial funding from ERC Grant No. 306719, ANR MoS2ValleyControl, Programme Investissements d'Avenir ANR-11-IDEX-0002-02, reference ANR-10-LABX-0037-NEXT, RFBR, RF President grants MD-5726.2015.2 and NSh-1085.2014.2 and Dynasty Foundation -- ICFPM.

\end{document}